\documentclass[prl,twocolumn,aps,floats,showpacs]{revtex4}

\usepackage{graphicx}

\begin{document}

\title{Realization of a photonic CNOT gate sufficient for quantum computation}

\author{Sara Gasparoni, Jian-Wei Pan\protect\footnote{Present address: Physikalisches Institut, Universit\"{a}t
Heidelberg, D-69120 Heidelberg, Germany}, Philip Walther, Terry
Rudolph\protect\footnote{Present address: Imperial College,
Blackett Labs, Prince Consort Rd, London SW7 2AZ} \& and Anton
Zeilinger\protect\footnote{also at Institut f\"{u}r Quantenoptik
und Quanteninformation, \"{O}sterreichische Akademie
 der Wissenschaften}
}


\affiliation{Institut f\"{u}r Experimentalphysik, Universit\"{a}t
Wien \\ Boltzmanngasse 5, 1090 Wien, Austria}

\begin{abstract}
\noindent \textbf{We report the first experimental demonstration
of a quantum controlled-NOT gate for different photons, which is
classically feed-forwardable. In the experiment, we achieved this
goal with the use only of linear optics, an entangled ancillary
pair of photons and post-selection. The techniques developed in
our experiment will be of significant importance for quantum
information processing with linear optics.}
\end{abstract}

\maketitle

Polarization-encoded qubits are well suited for information
transmission in Quantum Information Processing
(QIP)\cite{bouwmeester00a}. In recent years, the polarization
state of single photons has been used to experimentally
demonstrate quantum dense coding \cite{mattle96}, quantum
teleportation \cite{dik97} and quantum cryptography
\cite{thomas00,naik00,gisin01}. However, due to the difficulty of
achieving quantum logic operations between independent photons,
the application of photon states has been limited primarily to the
field of quantum communication. More precisely, the two-qubit
gates suitable for quantum computation generically require strong
interactions between individual photons, implying the need for
massive, reversible non-linearities well beyond those presently
available for photons, as opposed to other physical systems
\cite{Schmidtkaler03}.

Remarkably, Knill, Laflamme and Milburn \cite{knill01} found a way
to circumvent this problem and implement efficient quantum
computation using only linear optics, photo-detectors and
single-photon sources. In effect they showed that
\emph{measurement induced nonlinearity} was sufficient to obtain
efficient quantum computation.

The logic schemes KLM proposed were not, however, economical in
their use of optical components or ancillary photons. Various
groups have been working on reducing the complexity of these gates
while improving their theoretical efficiency (see e.g. Koashi et
al. \cite{koashi01}). In an exciting recent development, Nielsen
\cite{nielsen04} has shown that efficient linear optical quantum
computation is in fact possible without the elaborate
teleportation and Z-measurement error correction steps in KLM.
This is achieved by creation of linear optical versions of
Raussendorf and Briegel's \cite{raussendorf} cluster states.
Nielsen's method works for any non-trivial linear optical gate
which succeeds with finite probability, but which, when it fails,
effects a measurement in the computational basis.

A crucial requirement of both KLM's and Nielsen's constructions is
\emph{classical feed-forwardability}. Specifically, it must be in
principle possible to detect when the gate has succeeded by
measurement of ancilla photons in some appropriate state. This
information can then be fed-forward in such a way as to condition
future operations on the photon modes.

Recently \cite{fran03,sanaka02,obrien03} destructive linear
optical gate operations have been realized. As they necessarily
destroy the output state, such schemes are not classically
feed-forwardable. In the present paper we report the first
realization of a CNOT gate which operates on two polarization
qubits carried by independent photons and that satisfies the
feed-forwardability criterion. Moreover, when combined with single
qubit Hadamard rotations to perform a controlled-sign gate (so as
to build the cluster states of \cite{raussendorf} via Nielsen's
method) this gate also satisfies the criterion that when it fails
the qubits can be projected out in the computational basis.

\begin{figure}[ht]
\begin{center}
\includegraphics[width=0.7\columnwidth]{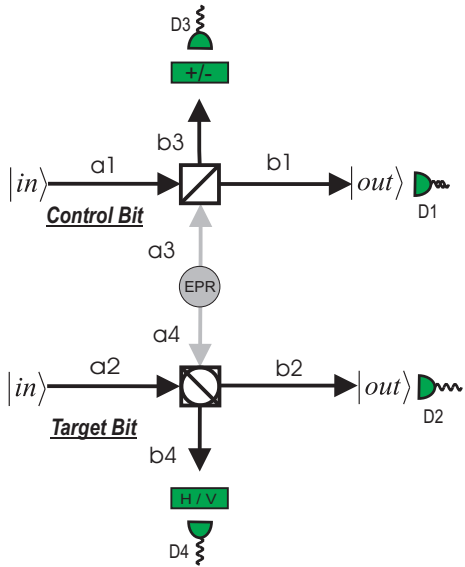}
\end{center}
 \caption{The scheme to obtain a photonic realization
of a CNOT gate with two independent qubits. The qubits are
encoded in the polarization of the photons. The scheme makes use
of linear optical components, polarization entanglement and
postselection. When one and only one photon is detected at the
polarization sensitive detectors in the spatial modes $b_3$ and
$b_4$ and in the polarization H, the scheme works as a CNOT
gate.} \label{cnot1}
\end{figure}

A CNOT gate flips the second (target) bit if and only if the first
one (control) has the logical value 1 and the control bit remains
unaffected. The scheme we  use to achieve the CNOT gate was first
proposed in \cite{fran01b} by Franson et al. and is shown in
Fig.\ref{cnot1}. This scheme performs a CNOT operation on the
input photons in spatial modes $a_1$ and $a_2$; the output qubits
are contained in spatial modes $b_1$ and $b_2$. The ancilla
photons in the spatial modes $a_3$ and $a_4$ are in the maximally
entangled Bell state

\begin{equation}
|\psi_{a_3a_4}\rangle=\frac{1}{\sqrt{2}}(|H\rangle_{a_3}|H\rangle_{a_4}+|V\rangle_{a_3}|V\rangle_{a_4})
\label{bellstate}
\end{equation}

In the following H (an horizontally polarized photon) and V (a
vertically polarized one) will denote our logical 0, 1. The scheme
works in those cases where one and only one photon is found in
each of the modes $b_3,b_4$; when both photons are H polarized no
further transformation is necessary on the output state (usually
this is referred to as passive operation). The scheme combines
two simpler gates, namely the destructive CNOT and the quantum
encoder. The first gate can be seen in the lower part of the
Fig.\ref{cnot1} and is constituted by a polarizing beam splitter
(PBS2) rotated by $45^\circ$ (the rotation is represented by the
circle drawn inside the symbol of the PBS), which works as a
destructive CNOT gate on the polarization qubits, as was
experimentally demonstrated in \cite{fran01b}. The upper part,
comprising the entangled state and the PBS1, is meant to encode
the control bit in the two channels $a_4$ and $b_1$. The photons
in the spatial modes $a_3$ and $a_4$ are in the maximally
entangled Bell state \ref{bellstate}

Thanks to the behaviour of our Polarizing Beam Splitter, that
transmits horizontally polarized photons and reflects vertically
polarized ones, the successful detection at the port $b_3$ of the
state $|+\rangle$ (the symbols $+,-$ stand for $H+V$ and $H-V$ )
post-selects the following transformation of the arbitrarily
input state in $a_1$

\begin{eqnarray*}
\alpha|H\rangle_{a_1}+\beta|V\rangle_{a_1}\rightarrow
\alpha|HH\rangle_{a_4b_1}+\beta|VV\rangle_{a_4b_1}
\end{eqnarray*}, thus we have the control bit encoded
in $a_4$ and in $b_1$, the photon in $a_4$ will be the control
input to the destructive CNOT gate, and will thus be destroyed,
while the second photon in b1 will be the output control qubit.

For the gate to work properly, we want the most general input
state
\begin{eqnarray*}
\vert\Psi_{a_1a_2}\rangle&=&\vert H\rangle_{a_1}(\alpha_1\vert
H\rangle_{a_2}+\alpha_2\vert V\rangle_{a_2})\\&+& \vert
V\rangle_{a_1}(\alpha_3\vert H\rangle_{a_2}+\alpha_4\vert
V\rangle_{a_2})
\end{eqnarray*}
to be converted to the output state
\begin{eqnarray*}
\vert\Psi_{a_1a_2}\rangle&=&\vert H\rangle_{a_1}(\alpha_1\vert
H\rangle_{a_2}+\alpha_2\vert V\rangle_{a_2})\\&+& \vert
V\rangle_{a_1}(\alpha_3\vert V\rangle_{a_2}+\alpha_4\vert
H\rangle_{a_2})
\end{eqnarray*}

 Let us
consider first the case where the control photon is in the
logical zero (H polarization state). The control photon will then
travel undisturbed through the PBS, arriving in the spatial mode
$b_1$. As required, the output photon is H polarized. In order
for the scheme to work a photon needs to arrive also at the
detector $D_3$ in $b_3$: given the input photon already in the
mode $b_1$, this additional photon comes necessarily from the EPR
pair, and is H polarized as it is transmitted by the PBS1. We
know that the photons in $a_3$ and $a_4$ are correlated
(\ref{bellstate}), so the photon in $a_4$ is also in the
horizontal polarization. Taking into account the $-45^\circ$
rotation of the polarization on the paths $a_2,a_4$ operated by
the half-wave plates, the input in the PBS2 will then be the
state $|--\rangle_{a_2a_4}(|-+\rangle_{a_2a_4})$ for a target
photon H (V) polarized. This state will give rise, with a
probability of $50\%$, to the state where two photons go through
the PBS2 $(|HH\rangle\pm|VV\rangle)_{b_2b_4}$ which, after the
additional rotation of the polarization,  and the subsequent
change to the H/V basis (where the measurement will be performed)
acquires the form
$(|HH\rangle+|VV\rangle)_{b_2b_4}((|HV\rangle+|VH\rangle)_{b_2b_4})$.
The expected result in the mode $b_2$ H(V) for the case where the
photon in $b_4$ is horizontally polarized. We can see in a
similar way that the gate works also for the cases where the
control photon is vertically polarized, or is polarized at
$45^\circ$.

\begin{figure}[ht]
\begin{center}
\includegraphics[width=0.7\columnwidth]{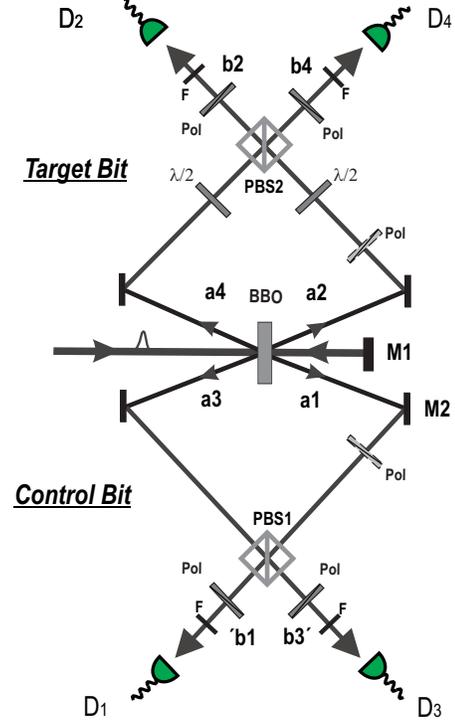}
\end{center}
 \caption{The experimental set-up. A type II Spontaneous
Parametric Down Conversion is used both to produce the ancilla
pair (in the spatial modes $a_3$ and $a_4$) and to produce the
two input qubits (in the spatial modes $a_1$ and $a_2$). In this
case initial entanglement polarization is not desired, and it is
destroyed by making the photons go through polarization filters
which prepare the required input state. Half-wave plates have
been placed in the photon paths in order to rotate the
polarization; compensators are able to nullify the birefringence
effects of the non-linear crystal and of the polarizing beam
splitters. Overlap of the wavepackets at the PBSs is assured
through spatial and spectral filtering. } \label{cnot2}
\end{figure}

Our experimental setup is shown in Fig.\ref{cnot2}. In order to
produce the entangled pair of ancilla photons in modes $a_3$ and
$a_4$, we use a type II Spontaneous Parametric Down Conversion
(SPDC) process; this pair is responsible for the transmission of
the quantum part of the information. We also need to produce the
two input qubits in the modes $a_1$ and $a_2$ to feed into the
gate. In our setup these input qubits are another SPDC pair, where
photon number entanglement is used and two photons are
simultaneously produced; the polarization entanglement is
destroyed by letting the photons pass through appropriate
polarizers. Thanks to these polarization filters, and to
appropriate half-wave plates, any desired two-qubits input state
can be prepared.

An ultraviolet pulsed laser, centered at a wavelength of 398nm,
with pulse duration 200fs and a repetition rate of 76MH, impinges
on a BBO crystal \cite{kwiat95a} producing probabilistically the
first pair in the spatial modes $a_1$ and $a_2$: these two
photons are fed into the gate as the input qubits. The UV laser
is then reflected back by the mirror M1 and, passing through the
crystal a second time, produces the entangled ancilla pair in
spatial modes $a_3$ and $a_4$. Half-wave plates and non-linear
crystals in the paths provide the necessary birefringence
compensation, and the same half-wave plates are used to adjust
the phase between the down converted photons (i.e. to produce the
state $\phi^+$) and to implement the CNOT gate.

We then superpose the two photons at Alice's (Bob's) side in the
modes $a_1,a_3$ $(a_2,a_4)$ at a polarizing beam splitter PBS1
(PBS2). Moving the mirror M1, mounted on a motorized translation
stage, allows to change the arrival time to make the photons as
indistinguishable as possible. A further degree of freedom is
afforded by the mirror M2, whose movement on a micrometrical
translation stage corrects slight asymmetries in the arms of the
set up. The indistinguishability between the overlapping photons
is improved by introducing narrow bandwidth (3nm) spectral
filters at the outputs of the PBSs and monitoring the outgoing
photons by fiber-coupled detectors. The single-mode fiber
couplers guarantee good spatial overlap of the detected photons;
the narrow bandwidth filters stretch the coherence time to about
700fs - substantially larger than the pump pulse duration
\cite{marek95}. The temporal and spatial filtering process
effectively erases any possibility of distinguishing the
photon-pairs and therefore leads to interference.

The scheme we have described allows the output photons to travel
freely in space, so that they may be further used in quantum
communication protocols, and this is achieved by detecting one
and only one photon in modes $b_3$ and $b_4$. The fact that we do
not yet have single-photon detectors for this wavelength at our
disposal actually forces us to implement a four-fold coincidence
detection to confirm that photons actually arrive in the output
modes $b_1$ and $b_2$.

So far, we have analyzed only the ideal case where exactly one
pair is produced at each passage of the UV light beam through the
BBO crystal. Let us see now how the two-pair events are going to
influence the outcomes. It is easy to see that our gate cuts out
those unwanted cases where two pairs are produced in the spatial
modes $a_3,a_4$: if the two photons in the spatial mode $a_3$ are
both horizontally or vertically polarized, either two or no
photons will arrive at the detector D3. The alternative case is
that the pairs in each spatial mode $(a_3,a_4)$ have two photons
that are in opposite polarization states; this could produce
either two or no photons  in paths $b_3$ and $b_4$. It is thus
proved that no four-fold event may occur from an event of this
kind.

Unluckily, this kind of noise is present for superposed qubits;
however, it may be easily measured and eliminated covering the
spatial modes $a_3$ and $a_4$ and measuring the four-fold
coincidences. Anyway, we note that the noise is not intrinsic in
the setup and is only due to practical drawbacks. Indeed, an
unbalancing method like the one used in \cite{pan03a} would allow
one to increase the signal to noise ratio to any desired value.

One more detail that should be addressed here is the problem of
birefringence. Each PBS introduces a small shift between the H and
V components, thus deteriorating the overlap of the photon
wavepackets.
 This
birefringence is responsible for the presence of unwanted terms
at the output state. The non-linear crystals put on the optical
path are able to compensate for this as well, as remarked
elsewhere \cite{pan03b}.

\begin{figure}[ht]
\begin{center}
\includegraphics[width=0.7\columnwidth]{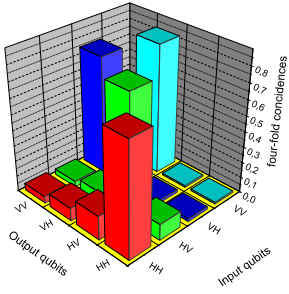}
\end{center}
 \caption{The above graph shows that the scheme works indeed for
the linear polarizations H,V. Four-fold coincidences for all the
possible (16) combinations of inputs and outputs are shown. When
the control qubit is in the logical value 0 (HH or HV), the gate
works as the identity gate. In contrast, when the control qubit
is in the logical value 1 (VH or VV), the gate works as a NOT
gate, flipping the second input bit. Noise is due to the non ideal
nature of the PBSs.} \label{cnot3}
\end{figure}

To experimentally demonstrate that the gate works, we first verify
that we obtain the desired CNOT (appropriately conditioned) for
the input qubits in states HH, HV, VH and VV. In Fig. \ref{cnot3}
we compare the count rates of all 16 possible combinations. We
see indeed that the gate is working properly in this basis.
Having verified this, we prove that the gate also works for a
superposition of states. The special case where the control input
is a $45^\circ$ polarized photon and the target qubit is a H
photon is very interesting: we expect that the state
$|H+V\rangle_{a_1}|H\rangle_{a_2}$ evolves into the maximally
entangled state $(|HH\rangle_{b_1b_2}+|VV\rangle_{b_1b_2})$. This
shows the reason why CNOT gates are so important: they can
transform separable states into entangled states and vice versa.
We input the state $|+\rangle_{a_1}|H\rangle_{a_2}$; first we
measure the count rates of the 4 combinations of the output
polarization (HH,..,VV) and observe that the contributions from
the terms HV and VH are negligible with a fidelity of $81\% $.

\begin{figure}[ht]
\begin{center}
\includegraphics[width=0.7\columnwidth]{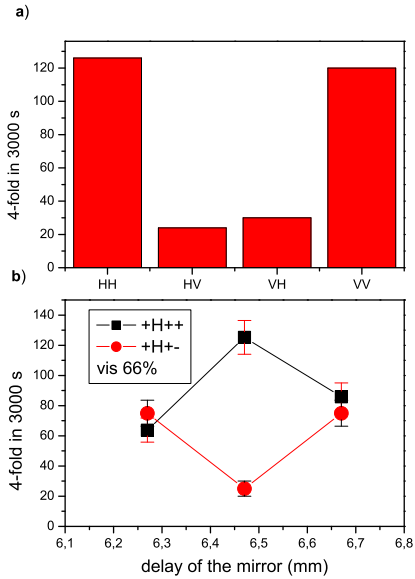}
\end{center}
\caption{Demonstration of the ability of the CNOT gate to
transform a separable state into an entangled state. In a) the
coincidence ratio between the different terms HH,..,VV is
measured, proving the birefringence of the PBS has been
sufficiently compensated, in b)the superposition between HH and
VV is proved to be coherent, by showing via Ou-Hong Mandel dip at
$45^\circ$ that the desired (H+V) state of the target bit emerges
much more often than the spurious state (H-V). The fidelity is of
$81\%\pm 2\%$ in the first case and $77\%\pm 3\%$ for the second.
} \label{cnot4}
\end{figure}

Then we prove that the output state is in a coherent
superposition, which is done by a further polarization
measurement. Going to the $|+\rangle,|-\rangle$ linear
polarization basis a Ou-Hong-Mandel interference measurement is
possible; this is shown in fig.\ref{cnot4}.

To sum up, the above demonstrated realization of a
feed-forwardable photonic CNOT gate uses only linear optics and
entanglement. The non-linearities required in such an interaction
are obtained through projective measurement of the ancilla pair.
Our result is an important progress in the direction of the
realization of a quantum computer. The price we pay for a
non-destructive scheme is the higher experimental sophistication,
particularly the necessity to use high precision timing and
coincidence techniques.

This work was supported  by the European Commission, contract
numbers ERBFMRXCT960087 and IST-1999-10033 and by the Austrian
Science Foundation (FWF), project number S6506.

\newpage

\bibliographystyle{nature}

   \end{document}